\documentclass[mathpazo]{cicp}%
\usepackage{amssymb}
\usepackage{color}
\usepackage{graphicx}
\usepackage{epstopdf}
\usepackage{amsmath}
\usepackage{verbatim}
\usepackage{amsfonts}%
\numberwithin{equation}{section}

\begin{document}

\title{Investigating the Selectivity of KcsA Channel by an Image Charge Solvation
Method (ICSM) in Molecular Dynamics Simulations}
\author{Katherine Baker\affil{1}, Duan Chen\affil{1}, and Wei Cai\affil{1}\comma\corrauth}

\begin{abstract}
In this paper, we study the selectivity of the potassium channel KcsA by a
recently developed image-charge solvation method(ICSM) combined with molecular
dynamics simulations. The hybrid solvation model in the ICSM is able to
demonstrate atomistically the function of the selectivity filter of the KcsA
channel when potassium and sodium ions are considered and their distributions
inside the filter are simulated. Our study also shows that the reaction field
effect, explicitly accounted for through image charge approximation in the
ICSM model, is necessary in reproducing the correct selectivity property of
the potassium channels.

\end{abstract}

\address{
\affilnum{1}\ Department of Mathematics and Statistics, University
of North Carolina at Charlotte, Charlotte, NC 28223, USA} \emails{ {\tt wcai@uncc.edu} (W.~Cai)}

\maketitle

\section{Introduction}

Ion channels are membrane-spanning proteins that form a pathway for the
movement of ions through the cell membrane and they play significant roles in
a wide variety of biological processes. Some examples of their many functions
include the control of secretion of hormones into the bloodstream, generating
electrical impulses that establish information transfer in the nervous system,
and controlling the pace of the heart and other muscles~\cite{UofIwebsite}.
The idea for assuming the existence of a means for transporting ions from the
exterior of a cell to the interior was proposed 63 years ago with Hodgkin and
Huxley\'{s} study of the electrical activity in squid giant
axon~\cite{Jordan-2005,Hodgkin-Huxley-1952}. They showed that both sodium and
potassium ions contributed to the ionic current and that their fluxes were in
opposite directions. Twenty years later Hladky and Haydon used small
antibiotic gramicidins to actually prove the existence of an ionic
pathway~\cite{Jordan-2005,Hladky-Haydon-1972}. Properties of ion channels and
their functions in manipulating electric currents by conducting different
ionic species heavily depend on their molecular structure in the presence of a
complicated surrounding solvent environment. In past decades, great technical
strides in many diverse areas of science culminated in the completion of x-ray
crystal structures of ion channels. Meanwhile, for theoretical studies, a
hierarchy of multi-scale mathematical models, from molecular
dynamics~\cite{AMBER, CHARMM22}, Brownian dynamics~\cite{Chung:2002}, and
Poisson-Nernst-Planck (PNP) theories~\cite{Chen:1997, Coalson:2005}, were
developed to study functions of ion channels.

\begin{figure}[th]
\begin{center}
\centering \includegraphics[scale=0.8]{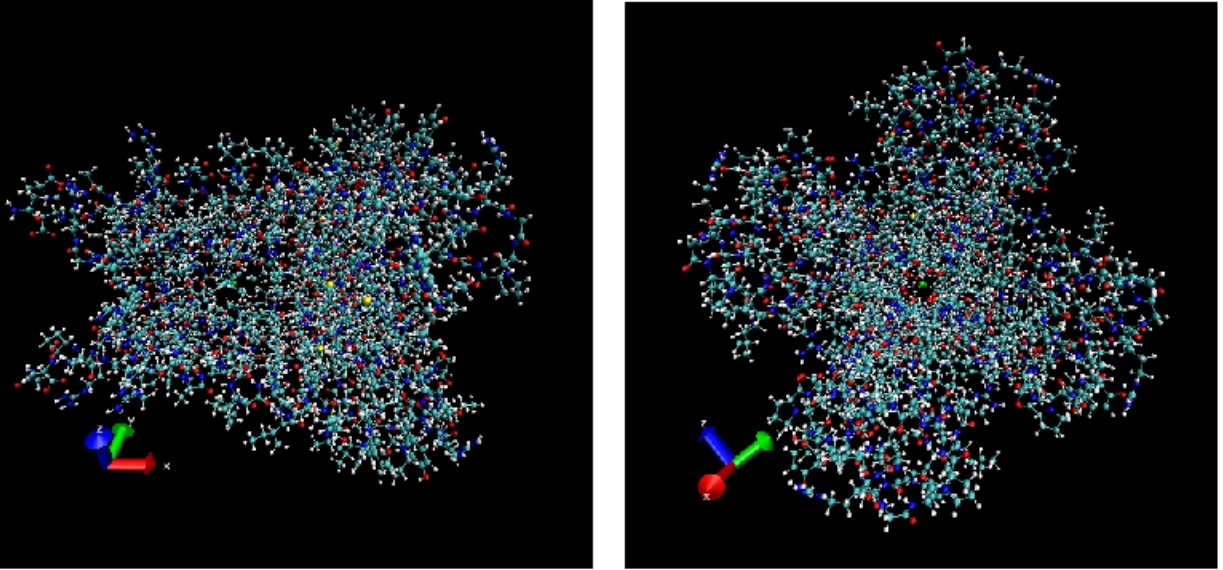}\newline%
\centering
\end{center}
\caption{An all-atom model of KcsA from a) Side view and b) extracellular end
view.}%
\label{figallatom}%
\end{figure}

\begin{figure}[th]
\begin{center}
\centering \includegraphics[scale=0.8]{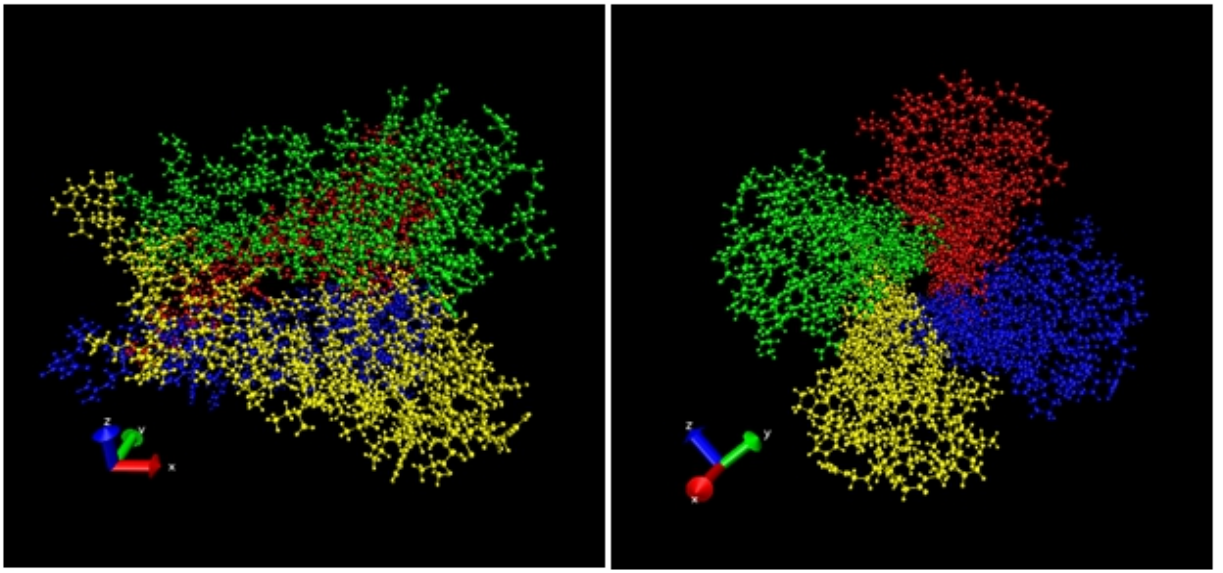}\newline%
\centering
\end{center}
\caption{View of the four identical subunits of the channel}%
\label{figproteins}%
\end{figure}

\begin{figure}[th]
\begin{center}
\centering \includegraphics[scale=0.8]{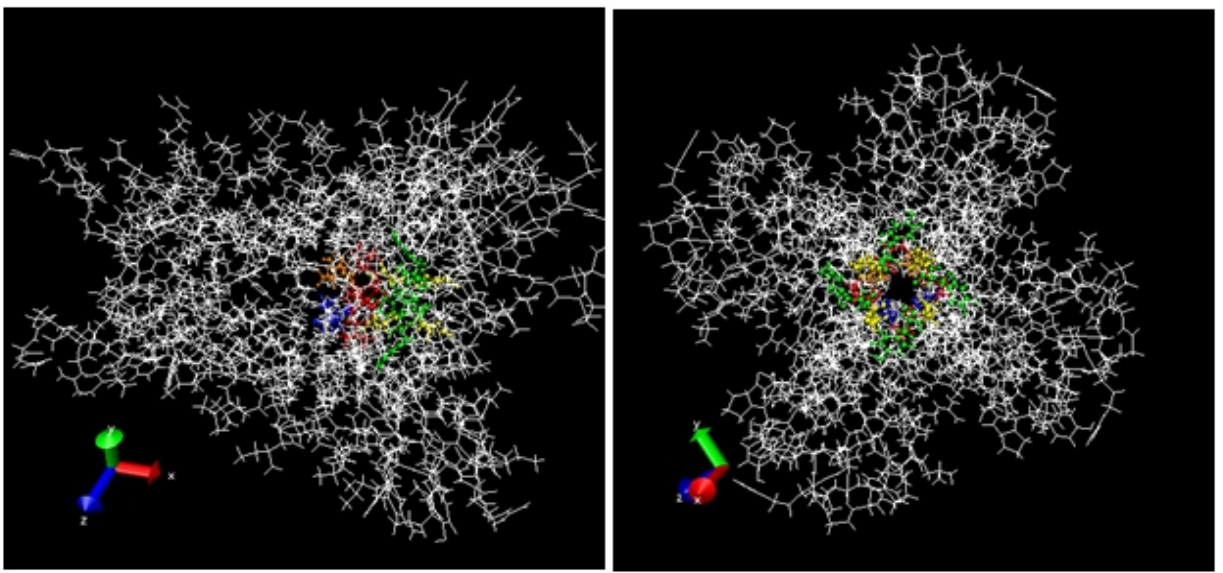}\newline\centering
\end{center}
\caption{The amino acids that make up the selection filter: blue --
threonine(T), red -- valine(V), yellow -- glycine(G), green -- tyrosine(Y).}%
\label{figaminos}%
\end{figure}

Potassium channels are specialized proteins able to facilitate and regulate
the conduction of ions, $\mathrm{K^{+}}$ ions in particular, through cell
membranes~\cite{UofIwebsite,Egwolf-Roux-2010}. In 1998, MacKinnon et
al.\cite{Jordan-2005,Hodgkin-Huxley-1952} successfully obtained the crystal
structure of KcsA (potassium crystallographically sited activation) channel at
a resolution of 2.0 \AA , allowing a direct laboratory observation of the
selectivity filter and binding sites of $\mathrm{K^{+}}$ ions. KcsA is
comprised of around 560 residues (Fig.~\ref{figallatom}) which form four
identical subunits (Fig.~\ref{figproteins}), each containing two alpha-helices
connected by a loop of approximately 30 amino acids. These proteins combine to
form three primary sections of the channel: the opening pore on the
cytoplasmic side of the cell interior, a small cavity ({of 5\AA ~ in radius})
filled with water and a mix of sodium ($\mathrm{Na^{+}}$) and potassium
($\mathrm{K^{+}}$) ions, and the selectivity filter. The selectivity filter,
{could be as narrow as 2\AA ~ in radius}, comprised of four specific cation
binding sites and formed by the backbone carbonyl groups of conserved residues
threonine (T), valine (V), glycine (G), and tyrosine (Y), allows fast
conduction of $\mathrm{K^{+}}$ while being highly selective for potassium ions
over sodium ions (Fig.~\ref{figaminos})~\cite{Egwolf-Roux-2010}.

The relatively complete functioning components (gating, selectivity,
conductance) and available high resolution structure of KcsA channel makes it
the most interesting case attracting investigation in biological studies and
mathematical modeling. However, modeling the selectivity of KcsA channel is
extremely challenging due to the complicated ion-water-protein interactions.
Counterintuitively, the $\mathrm{Na^{+}}$ ion has the same ionic valence as
the $\mathrm{K^{+}}$ ion does but with a smaller ionic radius, nevertheless it
is the one that is rejected by the narrow selectivity filter. In
\cite{Doyle-1998}, it was suggested that the small diameter of the selectivity
filter required dehydration of the cations entering the filter. To compensate
for the cost of the dehydration, the carbonyl oxygen atoms from the amino
acids in the filter take the place of the water oxygen atoms. The relative
rigidity of the filter precludes this action in the case of the
$\mathrm{Na^{+}}$ ions with a smaller radius but stronger binding of water
shell. However, a later study suggests that the selectivity can be explained
by the fact that the smaller $\mathrm{Na^{+}}$ does not bind to the
$\mathrm{K^{+}}$ sites in a thermodynamically favorable way
\cite{Thompson2009}.

To understand the selectivity mechanism of the filter, molecular dynamics
(MD), being an explicit atomistic model, is a naturally suitable method to
model these characteristics but the computation remains very expensive even
with current computer powers due to the large number of degree of freedom and
the necessarily small time step ($10^{-15}$ seconds) versus the ion permeation
time scale ($10^{-6}$ seconds)~\cite{Chung:2002}. The mean-field theory, such
as the PNP model, has successfully enhanced the model efficiency and
reproduced the channel conductance with carefully calibrated parameters.
Further, the PNP theory was strengthened recently to adopt the capacity of
modeling selectivity, by taking into account the ion-size effect and nonlocal
dielectric property of the solvent \cite{Burger:2007, Liu:2011, Zhou:2011,
xied:2013, lub:2014, Gillespie:2003}. Therefore, a highly efficient model with
the ability to retain the ion-water-protein interactions is indispensable to
investigate the properties, especially the selectivity, of the KcsA channel.

In this paper, we will conduct a study of the selectivity filter in the KcsA
channel with our recently developed Image Charge Solvation Method (ICSM),
implemented in the open-source Tinker MD package~\cite{tinker}. The ICSM is a
hybrid explicit/implicit solvation model developed to accurately account for
the reaction field of the solute-solvent environment. In contrast to the
traditional full MD simulations \cite{Biggin-2001,Shirivastava02,Berneche00},
most of the solvent in the system outside a designated spherical region is
modeled as a dielectric continuum, while only a limited number of particles
(protein, water molecules, and ions) inside the sphere are given an atomistic
description. The reaction field effect on the permanent charges of the protein
and mobile ions, due to the solvent/membrane surroundings, are accounted for
by a multiple image charge approximation
\cite{cai-dieSphImg-2007,deng-ionic-CCP2007,deng-ionic-JCP2007}. Therefore, in
the ICSM method, the treatment of electrostatic interaction does not use the
periodic condition as in the EWALD approach where infinitely many artificial
periodic copies of a simulated system, the ion-channel filter in this case,
will occur. The effect of the dielectric exterior to the simulation sphere is
accounted for easily by the image-charge method. Moreover, there is no
requirement of system neutrality in the hybrid ICSM model, not like EWALD-sum
based MD simulations. The efficiency, robustness, and capability of the ICSM
have been tested for homogeneous water system and solvation of
ions~\cite{Lin-2009}.

Here in this paper, we will use this hybrid solvation model to investigate the
positioning of sodium ions and potassium ions inside the selectivity filter of
the KcsA channel, to evaluate its likelihood of conducting the ions, and thus
the selective functions of this potassium channel. In order to balance the
accuracy and efficiency, cell membrane and intra/extra cellular solvents are
modeled as a dielectric continuum, most of the permanent charges on the
channel proteins are assumed as rigid, while the molecular dynamics is applied
to the selectivity filter, as well as ions and waters in the cavity chamber.
First of all, the fundamental physical properties of the KcsA channel, the
electrostatic landscape is obtained by the ICSM and compared to existing
results in literatures; four binding sites for $\mathrm{K^{+}}$ are
identified. Then, different combinations of $\mathrm{Na^{+}}$ and
$\mathrm{K^{+}}$ in the selectivity filter and cavity chamber are tested to
show that the movement of the ion in the cavity in fact depends on which ions
are in the selectivity filter, which will in turn determine what ions will
eventually be transported through the whole channel. Along with these
investigations, the inclusion of the reaction field in the hybrid ICSM will
also be examined and will be shown to be indispensable for accurate portrayal
of the ion conduction process.

The rest of paper is organized as follows: Section \ref{sec:icsm} is a review
of the ICSM. Simulation results of the selectivity filter of the KcsA are
given in Section \ref{sec:results}, and the paper ends with concluding remarks
in Section \ref{sec:conclusion}.

\section{ICSM - Image Charge Solvation Method}

\label{sec:icsm}

\begin{figure}[th]
\begin{center}
\includegraphics[scale=0.7]{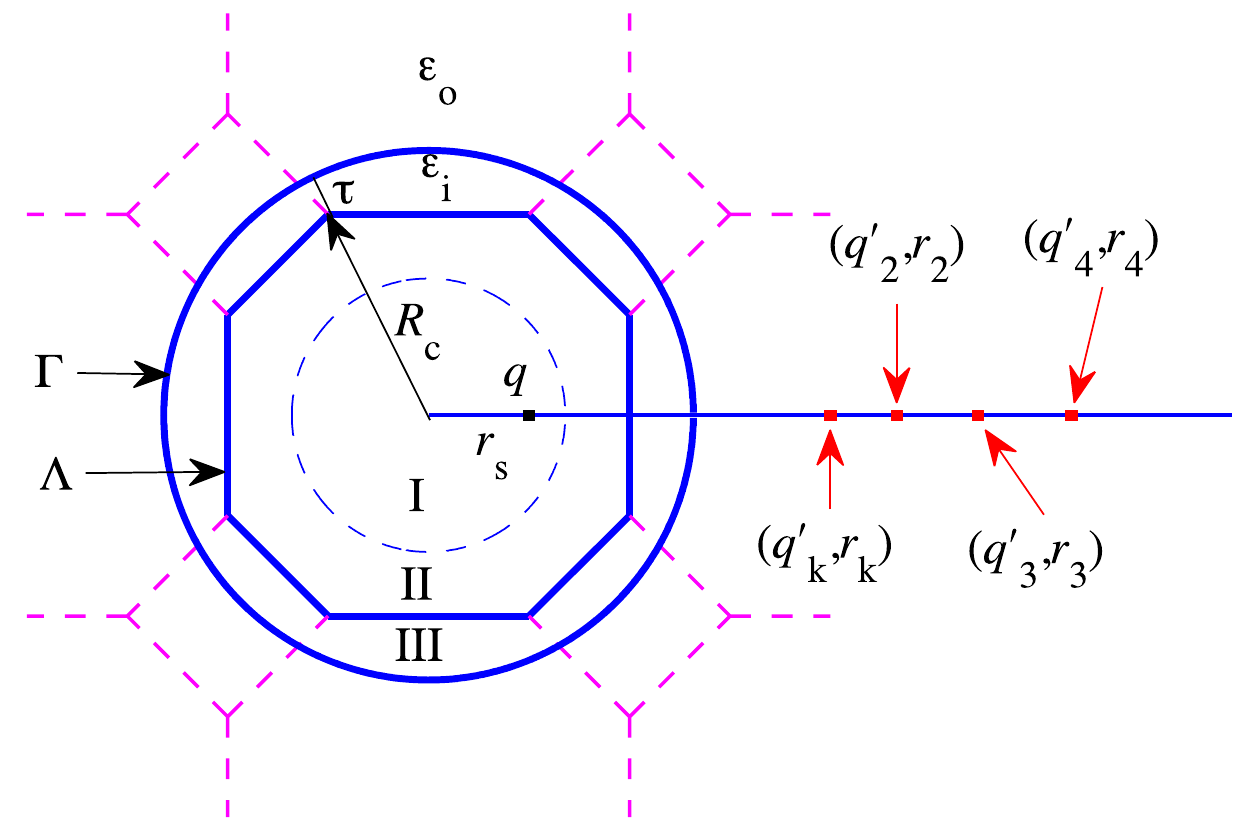}
\end{center}
\caption{A schematic of the ICSM system. Region I, indicated by a broken
circular line, contains the solute which can have a maximum diameter of
$d=(\sqrt{3}-\sqrt{5}/2)L-2\tau$. Particles in this area are not imaged.
Region II contains particles which have periodic images in Region III. Region
III contains the nearest periodic images of the particles in Region II. The
polarization of the solvent by the source charge $q$ at position
$\mathbf{r}_{s}$ results in the reaction field $\Phi_{RF}(r,\theta)$ that is
approximated by the potential created by the image charges, $q_{k}^{\prime}$
and $q_{i}^{\prime},i\geq2$, located at positions $r_{k},r_{i},i\geq2$.}%
\label{figicsmsystem}%
\end{figure}

The original setup of the ICSM is shown in a schematic of Figure
\ref{figicsmsystem}. A regular truncated octahedron (TO) with a size $R_{c}$
(the distance between its center and farthest corner) is employed as the main
simulation box. Inside a spherical domain (dashed circle with radius $a$)
labeled as region I, the solute molecule is placed in the central area of the
TO. Region II is the remainder of the TO box with Region I excluded, and this
region contains the solvent. Both the solute and solvent in regions I and II
are described explicitly. In order to reduce the possible surface effects
produced by the boundary of regions of atomistic and continuum descriptions, a
buffer layer of thickness $\tau$ is constructed by adding a larger spherical
domain containing the TO box. The area outside the TO box but inside the
larger sphere is denoted as region III and it is filled with periodic images
of the solvent molecules from Region II. Finally, the remaining solvent
outside the larger spherical cavity of radius $R_{c}+\tau$\ is modeled
implicitly as a dielectric continuum, whose reaction field effect on any
physical charge inside the larger sphere is approximated by multiple image
charges (located outside the larger sphere) and the later will be included in
the calculation of electrostatic interactions with the physical charges inside
the spherical cavity of radius $R_{c}+\tau$.

The multiple image charge approximation of the reaction field inside an
dielectric sphere extends the Kelvin image concept \cite{Jackson-1999} for a
conducting sphere. To simulate the solvation of a protein inside a dielectric
sphere containing water molecules, both the source charges and the field
points are inside the sphere. For each source charge, in addition to an image
point charge at the Kelvin image inversion point, there will be other image
charges distributed along a ray starting from the inversion point
\cite{cai-dieSphImg-2007} and a short summary of the multiple image charge
approximation of the reaction field is given below.

Given a local volume $\Gamma$ of a spherical shape of radius $R,R=R_{c}+\tau$
in our case, and dielectric permittivity $\varepsilon_{\mathrm{i}},$ embedded
in an infinite solvent of dielectric permittivity $\varepsilon_{\mathrm{o}}$,
the total electrostatic potential $\Phi(\mathbf{r})$ satisfies the Poisson and
Poisson-Boltzmann equation:
\begin{subequations}
\begin{align}
\varepsilon_{\mathrm{i}}\Delta\Phi(\mathbf{r})=-\rho_{in}(\mathbf{r}),\quad &
\mathrm{if\ \ }\mathbf{r}\in V_{\mathrm{in}},\label{poissona}\\
(\Delta-\lambda^{2})\Phi(\mathbf{r})=0,\quad &  \mathrm{if\ \ }\mathbf{r}\in
V_{\mathrm{out}},\label{poissonb}%
\end{align}
where the charge distribution inside the solute domain $V_{\mathrm{in}}$,
$\rho_{\mathrm{in}}(\mathbf{r})=\sum_{i}q_{i}\delta(\mathbf{r}-\mathbf{r}%
_{i})$ contains all explicit charges of the solute and solvent molecules. And
for the implicit region $V_{\mathrm{out}}$, $\lambda$ is the inverse
Debye-H\"{u}ckel screening length \cite{Lin-2009} and $\lambda=0$ if the
medium is considered as simple dielectric with no mobile ion density as in
this work. Here $\delta(\cdot)$ denotes the Dirac delta function. The two
equations are accompanied with the following interface and boundary
conditions:
\begin{equation}
\begin{aligned} \Phi(\mathbf{r}^+)& =\Phi(\mathbf{r}^-)\\ \epsilon_i\nabla \Phi\cdot \vec{n}|_{\mathbf{r}^+}& =\epsilon_o\nabla \Phi\cdot \vec{n}|_{\mathbf{r}^-}\\ \Phi(\infty)& =0, \end{aligned}
\end{equation}
where $\mathbf{r^{+}}$ and $\mathbf{r^{+}}$ represent the limiting value of
$\mathbf{r}$ when approaching to the interface $\Gamma$ from two sides, and
$\vec{n}$ is the unit normal direction vector.

Due to the principle of superposition, we only need to consider the case of
one single source charge in $V_{\mathrm{in}}$, i.e. $\rho_{\mathrm{in}%
}(\mathbf{r})=q\delta(\mathbf{r}-\mathbf{r}_{s})$, and the total potential can
be written as $\Phi=\Phi_{\mathrm{S}}+\Phi_{\mathrm{RF}}$ where $\Phi
_{\mathrm{RF}}$ is the primary field that results from the source charge $q$
at $r_{s}$ and $\Phi_{\mathrm{RF}}$ is the reaction field from the exterior
dielectric medium with dielectric constant $\varepsilon_{\mathrm{o}}$ and the
inverse Debye-H\"{u}ckel screening length $\lambda=0$, respectively
\cite{Baker-2013, Lin-2009}. The reaction field can be approximated by a set
of discrete image charges based on Gauss-Radau quadratures as
\cite{cai-dieSphImg-2007}:%

\end{subequations}
\begin{equation}
\Phi_{\mathrm{RF}}(\mathbf{r})\approx\frac{{q}_{\mathrm{k}}^{\prime}}%
{4\pi\varepsilon_{\mathrm{i}}|\mathbf{r}-\mathbf{r}_{\mathrm{k}}|}%
+\sum\limits_{m=2}^{N_{i}}\frac{q_{m}^{\prime}}{4\pi\varepsilon_{\mathrm{i}%
}|\mathbf{r}-\mathbf{r}_{m}^{\prime}|}, \label{phi}%
\end{equation}
where the subscript $\mathrm{k}$ is for the Kelvin image and $m\geq2$
represents the remainder of the discrete image charges, ${q}_{\mathrm{k}%
}^{\prime}={q}_{\mathrm{k}}+q_{1}^{\prime}=(1+\omega_{1}\varepsilon
_{\mathrm{i}}/2\varepsilon_{\mathrm{o}}){q}_{\mathrm{k}}$, ${q}_{\mathrm{k}%
}=\gamma\frac{R}{r_{s}}q,\gamma=\frac{\varepsilon_{\mathrm{i}}-\varepsilon
_{\mathrm{o}}}{\varepsilon_{\mathrm{i}}+\varepsilon_{\mathrm{o}}}$ and for all
$m\geq1$%

\begin{equation}
\ q_{m}^{\prime}=\frac{\varepsilon_{\mathrm{i}}(\varepsilon_{\mathrm{i}%
}-\varepsilon_{\mathrm{o}})}{2\varepsilon_{\mathrm{o}}(\varepsilon
_{\mathrm{i}}+\varepsilon_{\mathrm{o}})}\frac{\omega_{m}r_{m}}{R}q,\quad
r_{m}=r_{\mathrm{k}}\left(  \frac{2}{1-s_{m}}\right)  ^{1+\varepsilon
_{\mathrm{i}}/\varepsilon_{\mathrm{o}}}.
\end{equation}
Here $\{s_{m},\omega_{m}\}_{m=1}^{N_{i}}$ represent the points and the weights
of the Gauss-Radau quadrature \cite{gautschi-ORTHPOL-1994}. Image
approximation can also be derived for ionic solvent media
\cite{deng-ionic-JCP2007} \cite{Xucai:2009}\cite{Xuz:2015}.

\section{Simulation Results}

\label{sec:results}

In the present study, we use the KcsA channel with the PDB ID 2A9H
\cite{Yu-2005}. The data from the PDB was converted to a Tinker input file
using the built-in program pdbxyz.x, with the associated toxin removed and
five water molecules added to the central cavity.

\subsection{Simulation setup for the selectivity filter}

\begin{figure}[th]
\begin{center}
\centering \includegraphics[scale=0.6]{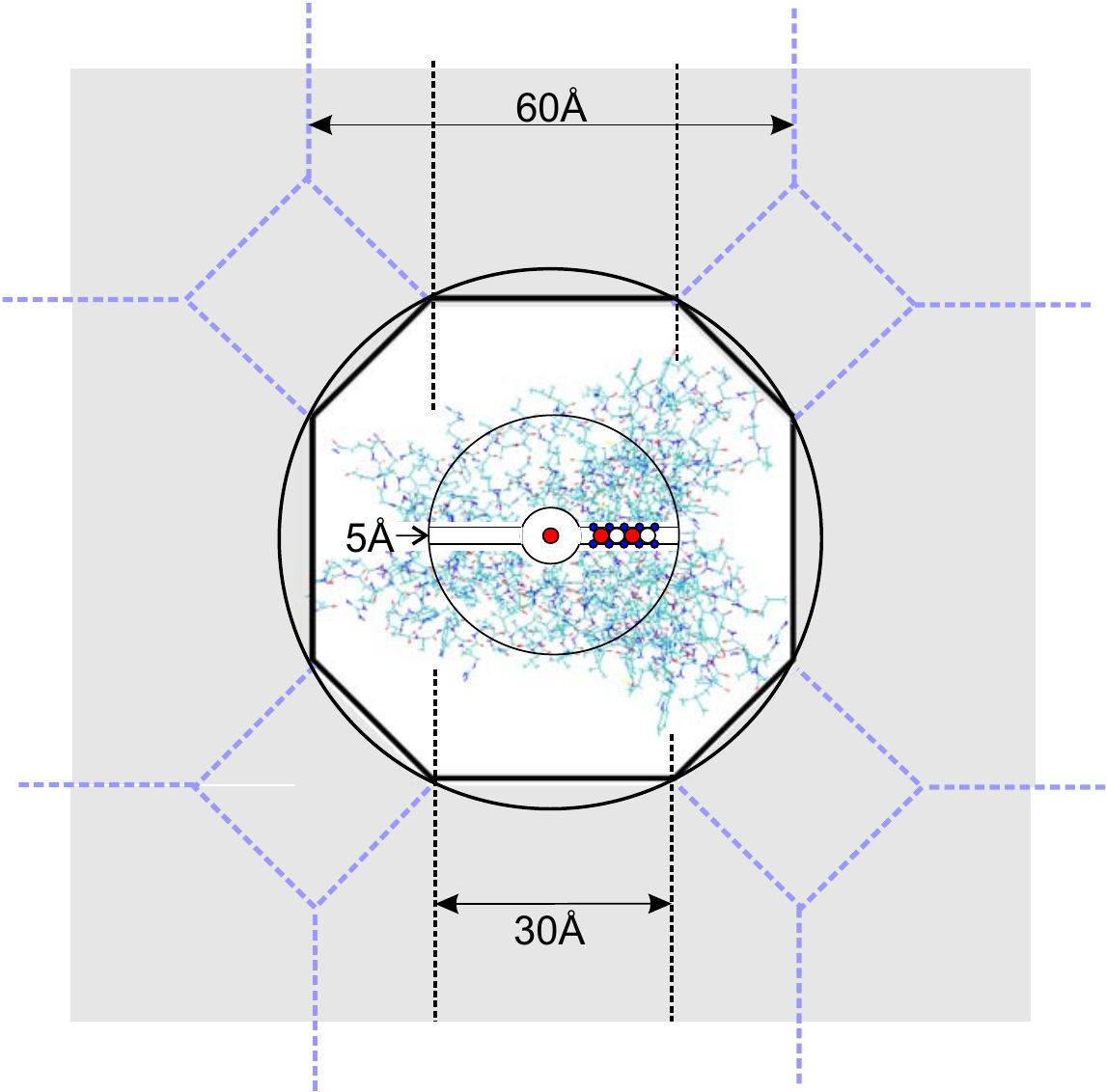}\newline%
\centering
\end{center}
\caption{A schematic showing the setup for the KcsA system. The red dots are
ions. The gray area is the dielectric continuum with $\varepsilon_{o}=2$ to 80
, the area in the circle of a diameter 30 \AA ~ is the productive region. The
dielectric inside the TO box is $\varepsilon_{i}=1$. The area between the
dashed black lines represents the size of the cellular membrane.}%
\label{figsyswithprot}%
\end{figure}

The ICSM is modified to simulate the selectivity filter inside the KcsA
channel. The channel pore of the KcsA, including the cavity, selectivity
filter, and residential ions are of greatest interest, so they are placed
inside Region I in the ICSM schematic of Fig. 4, along with a small portion of
the protein structure (see Fig.\ref{figsyswithprot}). Meanwhile, most of the
channel protein, some of the cell membrane and surrounding solvent are located
in Region II. The dielectric permittivity inside the TO box is denoted as
$\varepsilon_{\mathrm{i}}$ while $\varepsilon_{\mathrm{o}}$ is the dielectric
constant outside the larger sphere of radius $R_{c}+\tau$. As the difference
between the two dielectric constants (and therefore the surface effect) is
small, the buffer layer thickness $\tau$ is set to be zero.

\begin{figure}[th]
\begin{center}
\centering \includegraphics[scale=.4]{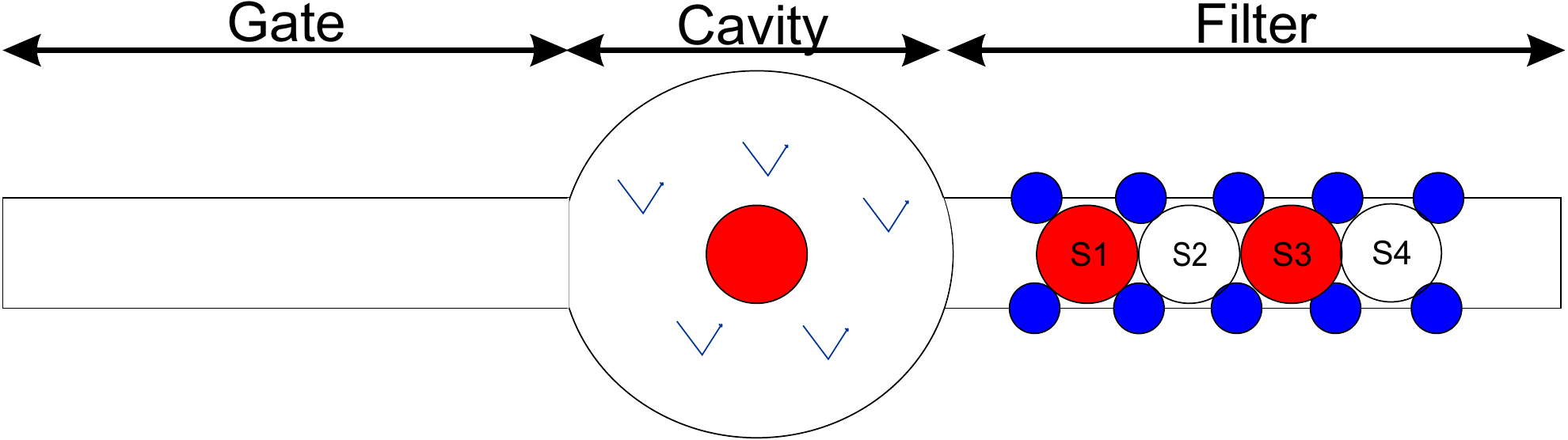}\newline%
\centering
\end{center}
\caption{Example of channel with three ions. One ion typically remains in the
water-filled cavity and two in the filter area. The blue dots represent the
carbonyl oxygen from the amino acids that make up the selectivity filter.}%
\label{figschemchannel}%
\end{figure}

Figure \ref{figschemchannel} gives a zoomed-in view of the channel pore of the
KcsA. The channel is set to run along the x-axis and its total length is taken
to be 60\AA . It contains three regions: the channel gate, a cavity chamber,
and a selectivity filter. In our coordinates, the channel starts from -30\AA ~
and ends at 30\AA ~. The positions of four standard binding sites for the ions
in the selectivity filter (labeled S1 - S4), which are located between the
carbonyl oxygen of the amino acids that make up the filter (shown in blue).

%We will
%consider the protein and the cell membrane as the exterior dielectric media
%outside the spherical cavity in the ICSM setup. The ions and waters inside the
%channel are considered the solute (Fig. \ref{figschemchannel}). Fig.
%\ref{figschemchannel} shows the . In our simulation, this channel is set to run along the
%x-axis.

\bigskip

The system is first minimized with all particles allowed to move freely using
the minimize program in Tinker and then run for 200ps for an initial
equilibrium. For subsequent simulations, we use the ACTIVE keyword in the key
file to lock down all atoms of the channel except the (sixteen) amino acids
making up the selection filter, the waters, and the ions. The parameters
contained in the Amber99 force field \cite{Amber-1981,Amber-2006}, also
included in Tinker, will be used for the MD simulations. The Velocity Verlet
algorithm was chosen for the time integration with a Nose-Hoover bath set at
300K. The time step is taken to be 2 fs, and the trajectory is recorded every
$0.1$ ps for analysis.

\subsection{Fundamental characteristics}

\label{sec:fundamental} Before the selectivity of the KcsA channel is
investigated, we use the ICSM to simulate some fundamental characteristics of
the channel, comparing with the results in the literature.

\begin{figure}[th]
\begin{center}
\centering \includegraphics[width=0.75\textwidth]{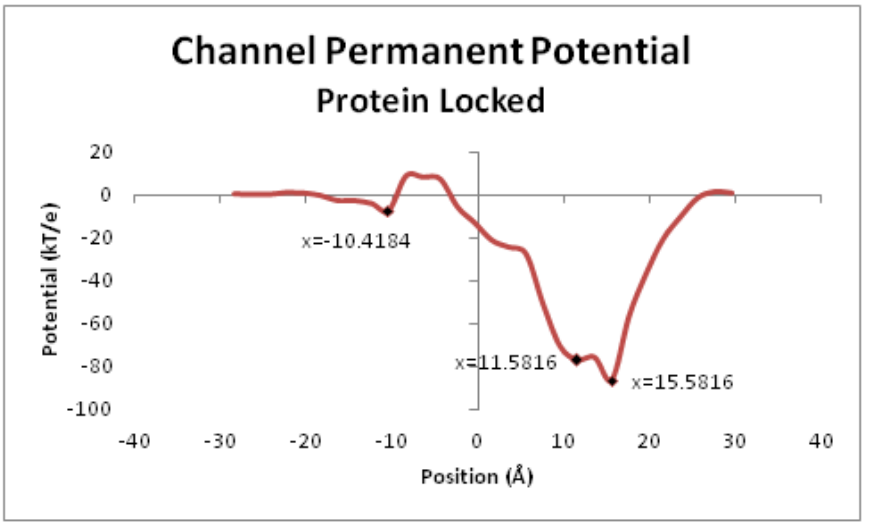}\newline%
\centering
\end{center}
\caption{Potential energy profile of KcsA prior to minimization.}%
\label{figpermpot}%
\end{figure}

\bigskip\textit{Case 1. Profile of electrostatic landscape of the channel: }
We first compute the channel permanent potential using the program analyze.x
of the Tinker package. This routine was modified to output the potential
energy acting on a particular ion as the latter moves through the channel
beginning at $x=-30$\AA and ending at $x=30$\AA . The ion was moved $2$\AA at
a time and the potential energy was calculated on the ion at that position.
The results for the channel prior to minimization and equilibration are shown
in Figure \ref{figpermpot}, which closely resembles the channel permanent
potential used by Jung, et al. for their studies on ERINP \cite{Yung-2009}. In
our calculation we found energy minima at -10.4 \AA , 11.5\AA , and 15.6\AA ,
respectively. Next, we allowed all atoms to move freely in the system and
minimized the system using the minimization program in Tinker. The final
configuration will be then used as the initial input for the subsequent
numerical simulations in this paper. The interval $[-5\mathring{A}%
,5\mathring{A}]$ in the channel is considered as the filter, while
$[6\mathring{A},10\mathring{A}]$, $[10\mathring{A},13\mathring{A}]$,
$[13\mathring{A},16\mathring{A}]$, and $[16\mathring{A},19\mathring{A}]$ are
regarded as site S1, S2, S3 and S4, respectively.

\begin{figure}[th]
\begin{center}
\includegraphics[width=\textwidth]{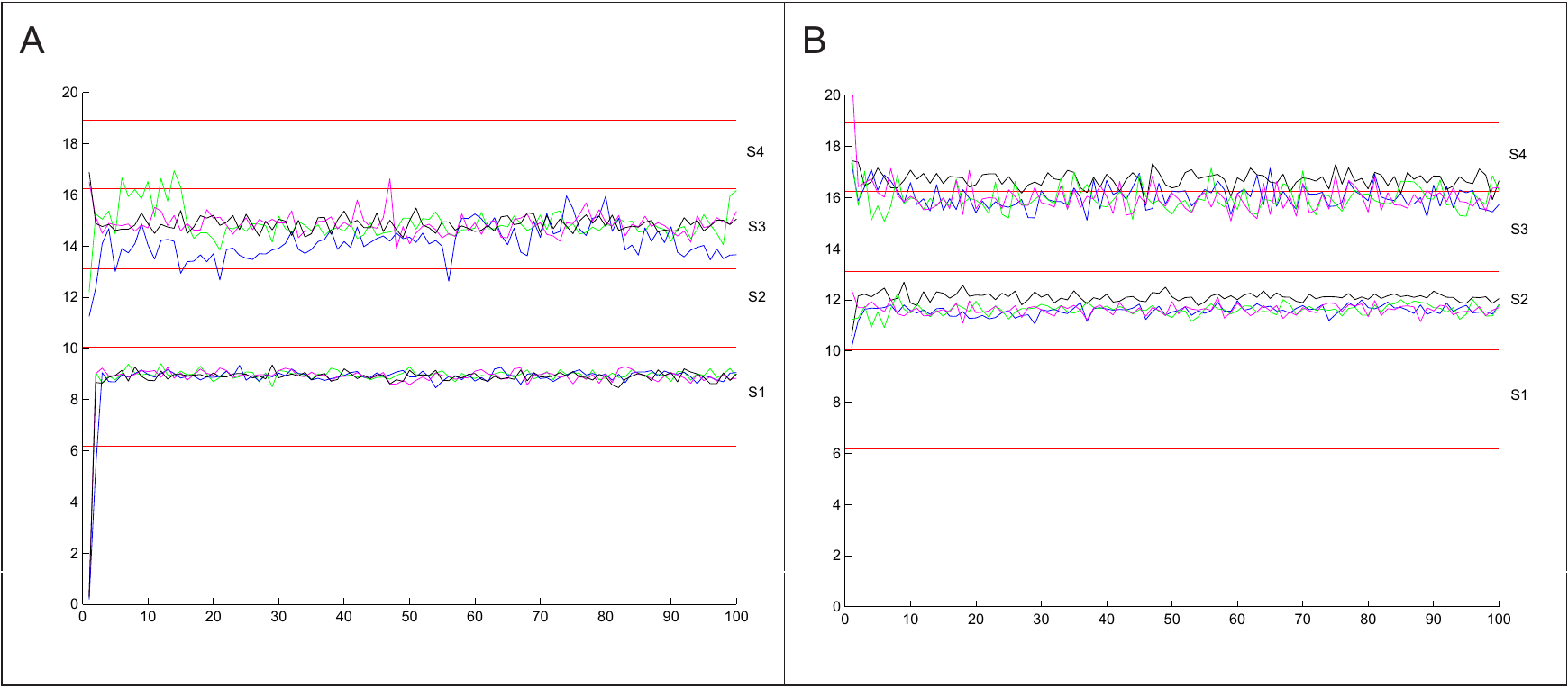}
\end{center}
\caption{Initial and final positions for tests with (A) - two potassium ions
in the channel and (B) - three potassium ions in the channel (the one in the
cavity is not plotted). The various colors represent different starting
positions. The sites S1-S4 are situated in between the five red lines, which
are corresponding to the position of filter residues. The unit for the
horizontal time axis is ps.}%
\label{2-3ions}%
\end{figure}

\bigskip\textit{Case 2.} \textit{Preferred positioning for two $\mathrm{K^{+}%
}$ inside the channel: }It was reported in \cite{Zhouyf:2001} that two
$\mathrm{K^{+}}$ ions would stay at site 1 (S1) and site 3 (S3) when a third
ion is far away. Otherwise they will occupy sites S2 and S4 when the third ion
is relatively close to the filter entrance. To verify this conclusion by the
ICSM, we run 100 ps simulations for several permutations of the positions of
only two potassium ions in the channel. One ion is located in the water cavity
(around $x=0.418$\AA ~) and the other is placed in different positions
(10-19\AA ~) in the selectivity filter. The detailed positions of the ions we
tested are summarized in Table \ref{table: table1}. The starting positions for
the ions in the filter were set at the midpoint between the carbonyl oxygen of
the filter amino acids. Regardless of their starting positions in the filter,
the two potassium ions come to rest at S1 (interval [6\AA , 11\AA ]) and S3
(interval [14\AA , 16\AA ]) , and the relaxation period for the ions is very
short. By approximately $3$ ps, the ions and waters in the filter reach their
approximate final positions. An example of the trajectories for four of these
runs is plotted in Figure \ref{2-3ions} (top). In this figure and all
remaining figures, the red horizontal lines indicate the position of the
oxygens for the Threonine (T), Valine (V), Glycine (G), Tyrosine (Y), and
Glycine (G). The intervals between TV, VG, GY, and YG are considered as S1,
S2, S3, and S4, respectively. In this instance, \textit{both of the ions move
into the filter and come to rest at positions S1 and S3}. And this
configuration will be a transitional phase between an ion exiting the channel
and another one entering when three ions are present in the channel for transport.

\bigskip\textit{Case 3.} \textit{Preferred positioning for three
$\mathrm{K^{+}}$ inside the channel: }In this case, we add a third potassium
ion into the filter (interval [10\AA , 16\AA ]) in addition to the two
$\mathrm{K^{+}}$ at the various places as in the previous study. The detailed
positions of the three ions are listed in Table \ref{table: table2}. After a
similar relaxation period, two of the ions rest at site S2 (interval [11\AA ,
14\AA ]) and site S4 (interval[16\AA , 19\AA ]), and another ion stays in the
water cavity but close to the entrance of the selectivity filter (around
$x=5$\AA ). Figure \ref{2-3ions} (bottom) shows the trajectories of the four
permutations of the positions of the three potassium ions listed in Table
\ref{table: table2}. In the final states, two ions stay between V-G and
between Y-G, respectively. The three ion configuration can be considered to be
a ``steady'' or normal state for the potassium channel and we note that
\textit{the final positions for the two ions in the selectivity filter are in
fact S2 and S4}, which are very close to the energy minima found in the empty
channel in Case 1 above.

\bigskip

\begin{table}[ptbh]
\caption{Initial and final positions for tests with two $K+$ ions in the
channel}%
\label{table: table1}
\begin{center}%
\begin{tabular}
[c]{|c|c|c|c|c|c|c|c|c|}\hline
& \multicolumn{2}{c|}{Position(\AA )} & \multicolumn{2}{c|}{Position(\AA )} &
\multicolumn{2}{c|}{Position(\AA )} & \multicolumn{2}{c|}{Position(\AA
)}\\\hline
& Initial & Final & Initial & Final & Initial & Final & Initial &
Final\\\hline
K1 & 0.418 & 8.635 & 0.418 & 9.049 & 0.418 & 8.957 & 0.418 & 8.923\\\hline
K2 & 18.544 & 14.739 & 15.800 & 14.717 & 13.324 & 15.387 & 10.347 &
13.451\\\hline
\end{tabular}
\end{center}
\end{table}

%The ion located in position S3 will then move in close proximity
%to the carbonyl oxygen belonging to Tyrosine(T). The starting and final positions for
%the examples shown can be seen in

\begin{table}[ptbh]
\caption{Initial and final positions for three ions in the channel}%
\label{table: table2}
\begin{center}%
\begin{tabular}
[c]{|c|c|c|c|c|c|c|c|c|}\hline
& \multicolumn{2}{c|}{Position(\AA )} & \multicolumn{2}{c|}{Position(\AA )} &
\multicolumn{2}{c|}{Position(\AA )} & \multicolumn{2}{c|}{Position(\AA
)}\\\hline
& Initial & Final & Initial & Final & Initial & Final & Initial &
Final\\\hline
K1 & 0.418 & 5.310 & 0.418 & 5.404 & 0.418 & 8.562 & 0.418 & 5.092\\\hline
K2 & 13.324 & 11.808 & 10.347 & 11.813 & 10.347 & 12.047 & 15.800 &
11.695\\\hline
K3 & 18.544 & 16.280 & 15.800 & 15.727 & 18.544 & 16.656 & 18.544 &
16.385\\\hline
\end{tabular}
\end{center}
\end{table}

\subsection{Selectivity and positioning of potassium versus sodium ions}

\label{sec:selectivity}

%Compared with atomistic simulations, simulation based on implicit continuum
%representation of ion channel lacks the ability to distinguish different type
%of ions with the same charge.
Our simulations in Section \ref{sec:fundamental} indicate that in the ``normal
state'' of three potassium ions in the channel, two of them are located at
site S2 and site S4, while the third one stays in the area around $x=5$\AA ,
which is at the entrance to the selectivity filter. This is consistent with
the results in \cite{Biggin-2001,Shirivastava02,Berneche00}, and this location
corresponds to the favorable location for a $\mathrm{K^{+}}$ ion formed by the
P helix dipoles \cite{Biggin-2001,Roux-1999}.

The KcsA channel dominantly conducts potassium over sodium ions, although the
two type of ions have the same charge. To investigate the ability of the ICSM
system to distinguish between $\mathrm{K^{+}}$ ions and $\mathrm{Na^{+}}$
ions, we repeat the three ion tests with sodium ions in the selectivity filter
instead of potassium ions, in order to simulate the preferred positionings of
$\mathrm{Na^{+}}$ ions and $\mathrm{Na^{+}}$/$\mathrm{K^{+}}$ mixtures.

\bigskip

\begin{figure}[th]
\begin{center}
\centering \includegraphics[width=0.8\textwidth]{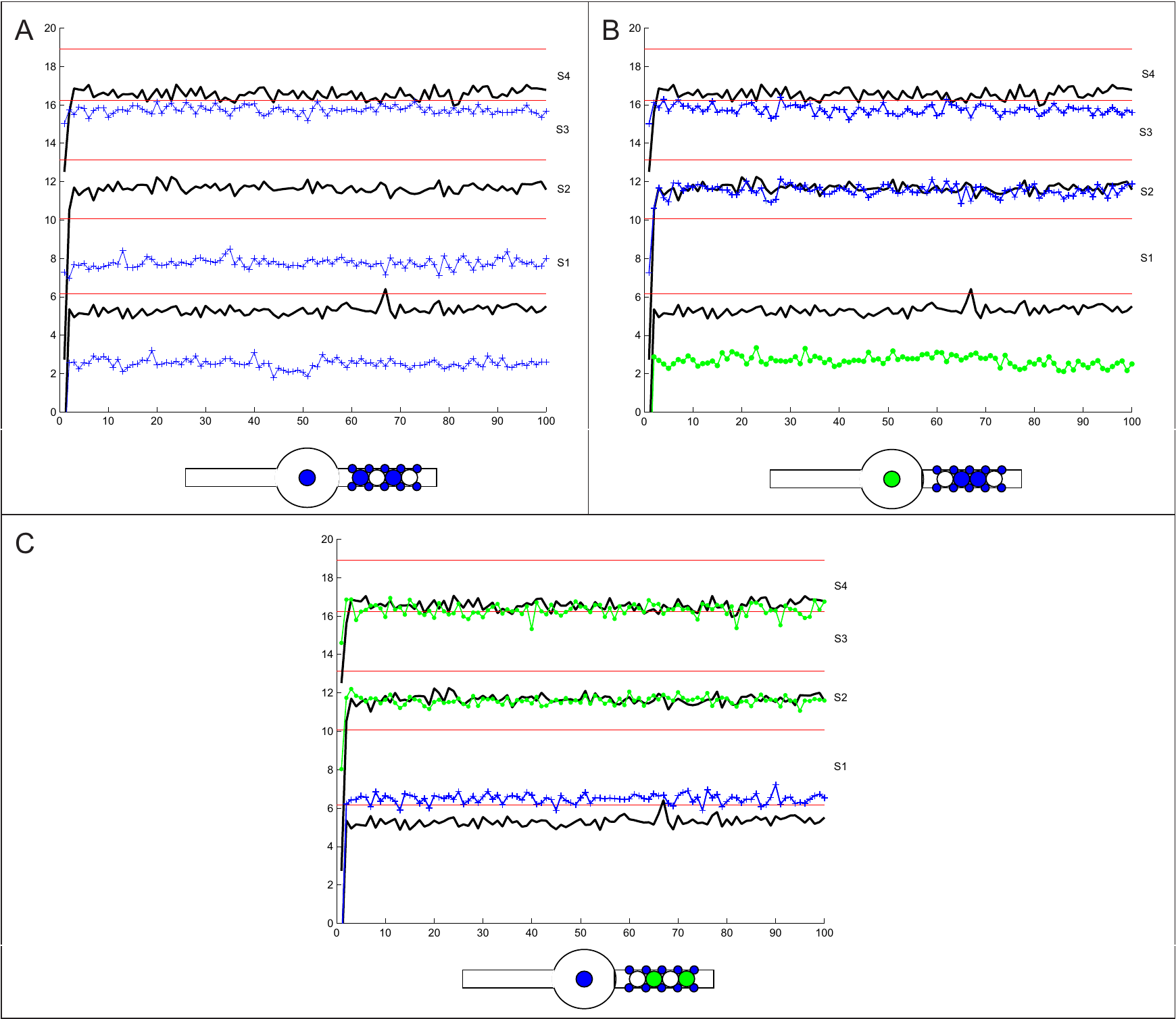}\newline%
\centering
\end{center}
\caption{Positioning of same ions in the selectivity filter. (A) two
$\mathrm{Na^{+}}$ in filter and a $\mathrm{Na^{+}}$ in the cavity; (B) two
$\mathrm{Na^{+}}$ in filter and a $\mathrm{K^{+}}$ in the cavity; (C) two
$\mathrm{K^{+}}$ in filter and a $\mathrm{Na^{+}}$ in the cavity. The black
solid lines represent case of 3 potassium ions for comparison, and the dotted
color lines are for a mix of potassium and sodium ions where green (circles)
is potassium and blue (+) is sodium. A schematic of the ion positions is given
under each figure. The unit for the horizontal time axis is ps.}%
\label{fignatestspure}%
\end{figure}

\textit{Case 1}: \textit{Preferred positioning of three $\mathrm{Na^{+}}$
inside the channel}: Figure \ref{fignatestspure}A shows the simulation results
if three $\mathrm{Na^{+}}$ ions are placed in the channel. In contrast to the
case of three-$\mathrm{K^{+}}$ (black curves), the two $\mathrm{Na^{+}}$ ions
(dashed grey) in the selectivity filter rest at the sites S1 and S3 (instead
of S2 and S4 as for the case of $\mathrm{K^{+}}$), while the third
$\mathrm{Na^{+}}$ ion stays around $x=2.5$\AA ~ in the cavity, which is toward
the center of the cavity and relatively away from the filter entrance. This
simulation suggests that when two $\mathrm{Na^{+}}$ are in the filter, they
are relatively away from the channel exit, meanwhile the third $\mathrm{Na^{+}%
}$ has difficulty approaching the filter entrance.

\bigskip

\medskip

\textit{Case 2}: \textit{One $\mathrm{K^{+}}$ in cavity and two
$\mathrm{Na^{+}}$ ions in filter}: With Case 1 in mind, we replace the
$\mathrm{Na^{+}}$ in the cavity by a $\mathrm{K^{+}}$ and keep the two
$\mathrm{Na^{+}}$ in the filter. From Fig. \ref{fignatestspure}B we can
clearly see that the $\mathrm{K^{+}}$ in the cavity pushes the $\mathrm{Na^{+}%
}$ at site S1 in Case 1 into site S2, hence the two $\mathrm{Na^{+}}$ ions in
the filter take sites S2 and S3. On the other hand, the $\mathrm{K^{+}}$
itself stays in the cavity around $x=2.5\mathring{A}$, which is relatively
away from the filter entrance. \medskip

%Figure \ref{fignatests}B is a test with two sodium ions in
%the selectivity filter and one potassium in the cavity. It clearly shows that
%sodium in the selectivity filter changes the positioning of a potassium ion in
%the cavity, keeping the ion closer to the center of the cavity. The sodium in
%the selectivity filter can also be seen to be in closer proximity to each
%other.
\bigskip\textit{Case 3. One $\mathrm{Na^{+}}$ in cavity and two $\mathrm{K^{+}%
}$ ions in filter}: As shown in Figure \ref{fignatestspure} C, in this case,
two potassium ions are already situated in the selectivity filter and a sodium
ion is located on the cellular side of the water-filled cavity. The two
$\mathrm{K^{+}}$ ions in the filter take the sites S2 and S4, as for the
three-$\mathrm{K^{+}}$ configuration. Comparing to the $\mathrm{K^{+}}$ in the
cavity in the three-$\mathrm{K^{+}}$ system, the position of the
$\mathrm{Na^{+}}$ ion is closer to the entrance of the filter and on the other
side of the carbonyl oxygen for the tyrosine. \medskip

\begin{figure}[th]
\begin{center}
\centering \includegraphics[width=0.8\textwidth]{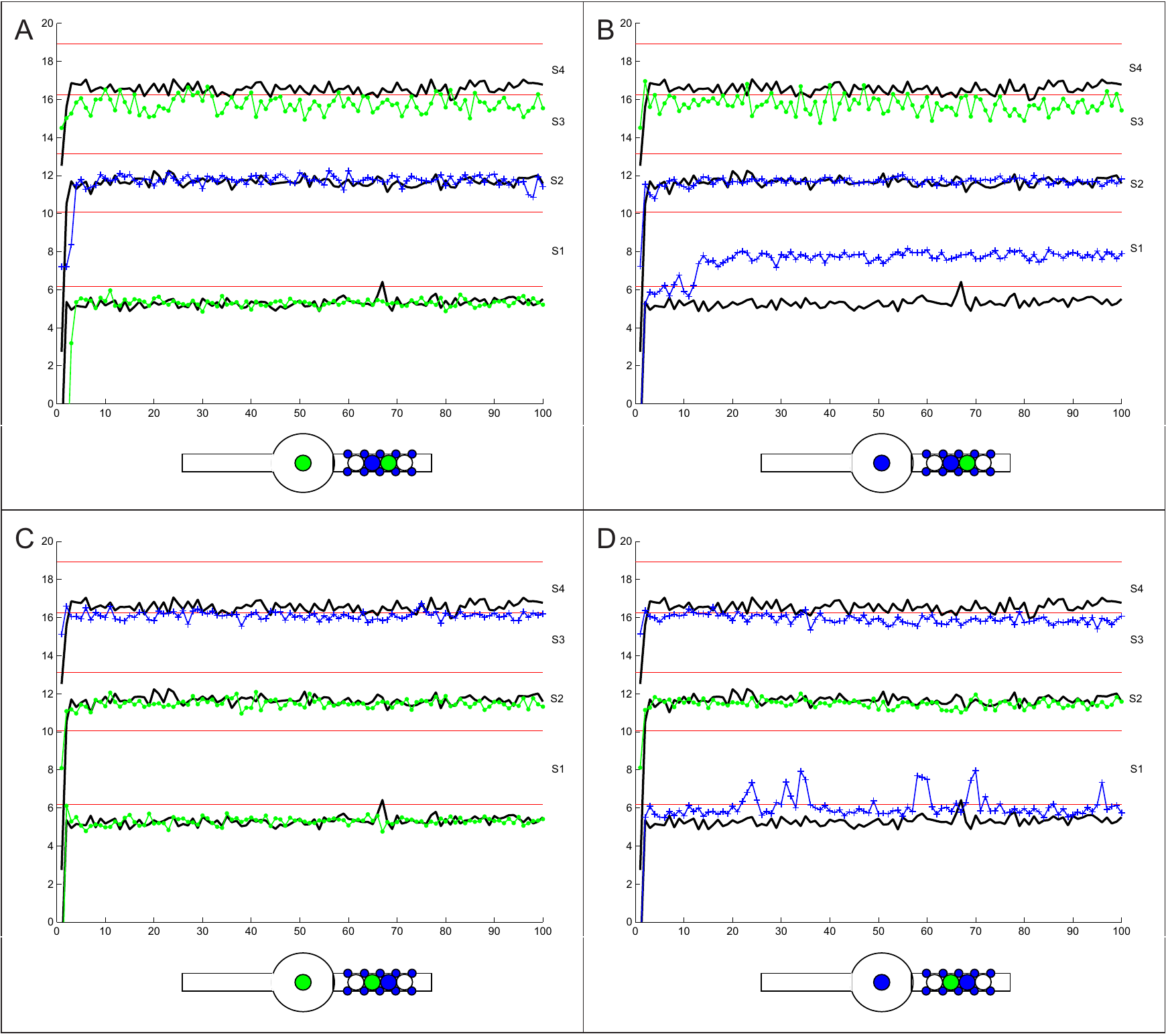}\newline%
\centering
\end{center}
\caption{Positioning of ion mixtures in the selectivity filter (from the
cavity to the filter exit direction). (A) $\mathrm{K^{+}Na^{+}K^{+}}$; (B)
$\mathrm{Na^{+}Na^{+}K^{+}}$; (C) $\mathrm{K^{+}K^{+}Na^{+}}$; (D)
$\mathrm{Na^{+}K^{+}Na^{+}}$. The black solid lines represent case of 3
potassium ions for comparison, and the dotted color lines are for a mix of
potassium and sodium ions where green (circles) is potassium and blue (+) is
sodium. A schematic of the ion positions is given under each figure. The unit
for the horizontal time axis is ps.}%
\label{fignatestsmixed}%
\end{figure}

\bigskip\textit{Case 4. $\mathrm{Na^{+}}$ and $\mathrm{K^{+}}$ mixtures in
filter}: Figure \ref{fignatestsmixed} shows the permutation of $\mathrm{Na^{+}%
}$ and $\mathrm{K^{+}}$ mixtures in the filter region, while in the cavity we
have either $\mathrm{Na^{+}}$ or $\mathrm{K^{+}}$. Figs. \ref{fignatestsmixed}%
A-D list the situations of $\mathrm{K^{+}Na^{+}K^{+}}$, $\mathrm{Na^{+}%
Na^{+}K^{+}}$, $\mathrm{K^{+}K^{+}Na^{+}}$, and $\mathrm{Na^{+}K^{+}Na^{+}}$,
respectively. The positions of the ions are ordered from the cavity to the
exit of the filter. It can be seen from the simulations that no ion can occupy
the site S4 in any of the ion combinations.\medskip

\medskip

\bigskip From the above simulations it can be summarized that in a
three-$\mathrm{Na^{+}}$ system, two sodium ions are occupying in the filter
region and stay at sites S1 and S3 positions, while the third ion in the
cavity is not able to move toward the proximity of the filter. The
$\mathrm{Na^{+}}$ at site S1 is able to proceed to site S2 if the sodium ion
in the cavity is replaced by a potassium ion but the other $\mathrm{Na^{+}}$
in the filter stays at site S3. Alternatively, when a $\mathrm{Na^{+}}$ in the
water cavity faces two $\mathrm{K^{+}}$ in the selectivity filter, it is
closer to the filter entrance than the position of the $\mathrm{K^{+}}$ in the
cavity in a three $\mathrm{K^{+}}$ system. Therefore, we conjecture that
$\mathrm{Na^{+}}$ ions are able to approach the filter entrance and
translocate from site S1 to S2, but no evidence shows that they are able to
move to site S4. This may be one reason for the reduced possibility of
$\mathrm{Na^{+}}$ ion conducting through the channel, consistent with the fact
that the KcsA channel is designed preferentially for three potassium ions in
the setting for potassium transporting through the channel as shown in Section 3.2.

\subsection{Effect of reaction field in ICSM}

We also investigated the importance of the long range reaction field by
running same configurations with the image charges and without image charges.
\medskip

\begin{figure}[th]
\begin{center}
\centering \includegraphics[width=0.8\textwidth]{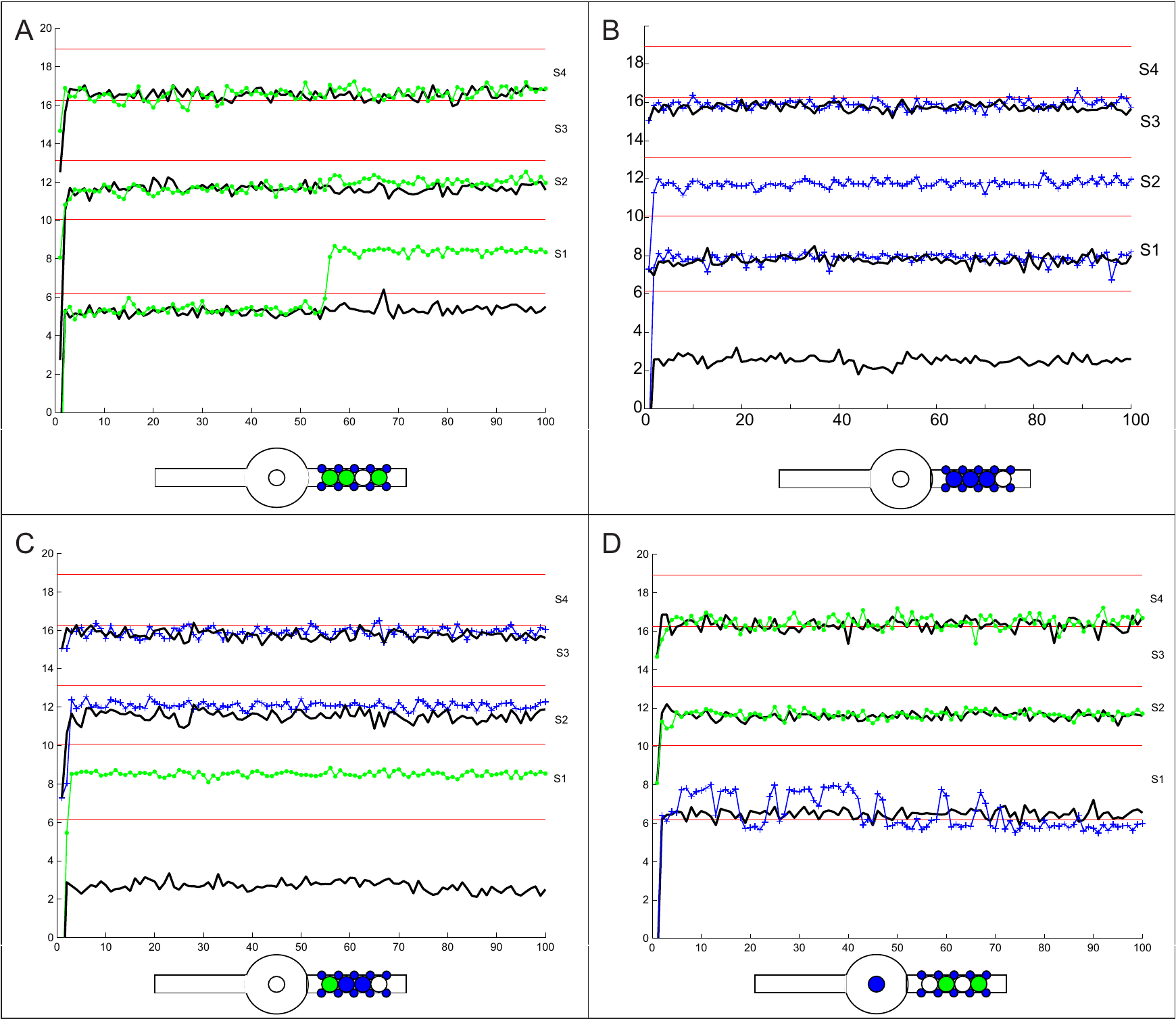}\newline%
\centering
\end{center}
\caption{Positioning of ion combinations in the channel (from the cavity to
the filter exit direction) without reaction field. (A) $\mathrm{K^{+}%
K^{+}K^{+}}$; (B) $\mathrm{Na^{+}Na^{+}Na^{+}}$; (C) $\mathrm{K^{+}%
Na^{+}Na^{+}}$; (D) $\mathrm{Na^{+}K^{+}K^{+}}$. The black solid lines
represent each case with the reaction field for comparison, and the dotted
color lines are for a mix of potassium and sodium ions where green (circles)
is potassium and blue (+) is sodium. A schematic of the ion positions is given
under each figure. The unit for the horizontal time axis is ps.}%
\label{figrftests}%
\end{figure}

\textit{Case 1. Simulation of the three-$\mathrm{K^{+}}$system without the
reaction field:} We first consider the stable configuration of three potassium
ions in the channel without taking into account the reaction field. For
comparisons, the corresponding simulation results with reaction field are also
provided. Fig. \ref{figrftests} A show that without the reaction field, the
two $\mathrm{K^{+}}$ ions in the selectivity filter are still occupying sites
S2 and S4. However, the final resting position for the third ion, which is
initially in the water cavity, is approximately at $x=8.3$\AA ~ when the
reaction field is neglected. In other words, this potassium ion is able to
move closer to entrance of the filter and eventually takes site S1(between the
residues T and V). Therefore, comparing to the results with reaction field for
the three-$\mathrm{K^{+}}$ system where two ions take sites S2 and S4 in the
filter while one stays in the water cavity at $x=5.5$\AA . Without the
reaction field the three potassium ions will take sites S1, S2, and S4; this
is against the current biological conclusions.

\medskip

\textit{Case 2. Simulation of the three-$\mathrm{Na^{+}}$ system without the
reaction field:} The positions of three $\mathrm{Na^{+}}$ ions in the channel
are plotted in Fig.\ref{figrftests}B when the simulation is implemented
without the reaction field. For a better comparison, the approximate final
positions of these ions are shown in Table \ref{table3}. Recall the
simulations with the reaction field, two sodium ions in the selectivity filter
take the sites S1 and S3, while the third ion stays in the cavity around
$x\approx2.6$\AA . But without the reaction field, all three sodium ions are
able to enter the filter region and they occupy the sites S1, S2, and S3,
respectively. Specifically, the ion in the cavity (sodium 1) moves toward the
filter and rests at $x\approx8.19$\AA ~ if the reaction field is absent.

\begin{table}[ptbh]
\caption{Comparison of the final positions for three sodium in the channel
with and without a reaction field.}%
\label{table3}
\begin{center}%
\begin{tabular}
[c]{|c|c|c|c|}\hline
& Sodium 1 & Sodium 2 & Sodium 3\\\hline
With RF & 2.59\AA  & 8.00\AA  & 15.66\AA \\\hline
No RF & 8.18\AA  & 11.99\AA  & 15.74\AA \\\hline
\end{tabular}
\end{center}
\end{table}

\medskip\textit{Case 3. Simulation of $\mathrm{K^{+}\slash Na^{+}}$ mixtures
in the channel without the reaction field:} For a further investigation of the
effect of the reaction field, the case of two $\mathrm{Na^{+}}$ in the
selectivity filter and a $\mathrm{K^{+}}$ in the cavity is considered. In
section \ref{sec:selectivity}, it is observed that the two $\mathrm{Na^{+}}$
in the filter take sites S2 and S3, and the $\mathrm{K^{+}}$ ion stays in the
cavity around the position $x\approx2.6$\AA . While ignoring the reaction
field, the $\mathrm{K^{+}}$ ion enters the filter and takes the position at
site S1. Thus, in this case, the selectivity filter will contain three
residential ions as $\mathrm{K^{+}}$ occupies S1 and the two $\mathrm{Na^{+}}$
ions occupy sites S2 and S3, as shown in Fig.\ref{figrftests}C. On the other
hand, for the case of two $\mathrm{K^{+}}$ in the filter and one
$\mathrm{Na^{+}}$ in the cavity, the results without the reaction field are
very close to the ones with the reaction field, however, the position of the
$\mathrm{Na^{+}}$ in the cavity fluctuates considerably more (see
Fig.\ref{figrftests}D).
%and it is possible that a
%longer timeframe might show it moving into the filter. \medskip

%\textit{Case 2.} \textit{Simulation of
%two $\rm Na^+$ in the filter and one $\rm K^+$ in the cavity without reaction field}: For graph B, there
%are two sodium in the selectivity filter and one potassium in the cavity. With
%the reaction field included, the potassium in the cavity stays close to the
%center of the cavity at 2.5\AA . Without the reaction field, this ion again
%moves into the selectivity filter to 8.5\AA .\medskip
%\textit{Case 3.} \textit{Effect of reaction field and} \textit{simulation of
%two potassium in the filter and one sodium in the cavity }: In Fig.
%\ref{figrftests}-C, there are two potassium in the selectivity filter and one
%sodium in the cavity. For this case,

From the above studies it can be summarized that the cavity ion, no matter
$\mathrm{K^{+}}$ or $\mathrm{Na^{+}}$, is able to move into the filter on the
other side of the threonine to position $x\approx8.2$\AA ~ if the reaction
field is neglected. This fact implies that there will be three ions in the
selectivity filter, and this does not agree with the biological observations.
Therefore, we can see clearly that the reaction field is critical for the
fidelity of the ICSM to reproduce the physical mechanism of the KcsA channel
in its selectivity and positioning of the ions. \medskip

\section{Conclusions}

\label{sec:conclusion}

In this paper, we have shown that the Image-Charge Solvation Method (ICSM) is
applicable to study the selectivity filter of a KcsA channel where only a
local spherical region around the filter section of the channel is modeled in
atomistic details and the effect of the rest of the channel and membrane and
solvents can be modeled as a continuum dielectric, whose reaction field is
approximated by a simple multiple image charge representation.

Our numerical simulations have produced the following results:

\begin{itemize}
\item The preferred positioning for two potassium ions inside the channel are
in sites \textit{ S1 and S3.}

\item The preferred positioning for three potassium ions inside the channel
are one in the cavity and two at sites \textit{S2} and \textit{S4}, which is
the characteristic positioning of three ions for conducting the potassium ions
by the KcsA channel.

\item The ICSM algorithm is able to distinguish sodium and potassium ions when
investigating the selectivity of the KcsA channel. In contrast to a
three-$\mathrm{K^{+}}$ system, the two $\mathrm{Na^{+}}$ ions in the
selectivity filter occupy sites \emph{S1} and \emph{S3} instead of \emph{S2}
and \emph{S4}, and the third $\mathrm{Na^{+}}$ in the water cavity stays
relatively away from the filter entrance. Among all the permutations of 3
ions, except for the case of 3 potassium ions, no ion was found to be able to
move to site \emph{S4, } this is consistent with the fact that the KcsA
channel is indeed a potassium channel designed to transport potassium ions.

\item The reaction field from the dielectric environment outside the atomistic
region containing the filter is critical in the accurate representation of the
selectivity function of the filter and the correct predictions of the ion
positioning inside the channel.
\end{itemize}

In future research, the following issues will be addressed: the layered
dielectric media outside the simulation region comprised of membrane and ionic
solvents as for the time being the latter is simply ignored; the I-V curve of
the KcsA channel calculated by ICSM molecular dynamics simulations.

\section*{Acknowledgement}

The authors acknowledge the support of the US Army Office of Research (Grant
No. W911NF-14-1-0297), the National Science Foundation (DMS-1315128), and the
NSFC (grant number 91230105) for the work in this paper. Authors also thank
the insightful comments from Drs. Xiaolin Cheng and Chun Liu on various
aspects of the simulations for KcsA channels.

\bibliographystyle{plain}
\bibliography{kcsaICSMkab}

\end{document}